\documentclass[epj]{webofc}
\usepackage[varg]{txfonts}   

\newcommand{\be}{\begin{equation}}  
\newcommand{\ee}{\end{equation}}  
\newcommand{\beq}{\begin{eqnarray}}  
\newcommand{\eeq}{\end{eqnarray}}

\newcommand{\Dlr}{\buildrel \leftrightarrow \over D\raise-1pt\hbox{}}
\woctitle{XLV International Symposium on Multiparticle Dynamics}
\begin{document}
\title{Selected results on hadron structure using state-of-the-art lattice QCD simulations}
%
%

\author{Constantia Alexandrou\inst{1,2}\fnsep\thanks{\email{alexand@ucy.ac.cy}} 
}

\institute{Department of Physics,
  University of Cyprus,
  P.O. Box 20537,
  1678 Nicosia,
  Cyprus
\and
           Computation-based Science and Technology Research Center,
  The Cyprus Institute,
  20 Kavafi Str.,
  Nicosia 2121,
  Cyprus 
          }

\abstract{%
 We review progress on hadron structure using lattice QCD simulations at or near to physical values of the QCD parameters. In particular, we discuss recent results  on  hadron masses, the nucleon charges, spin,  gluon and quark unpolarized moments, the axial charge of hyperons, and the pion unpolarized moment. 
}
\maketitle
\section{Introduction}
\label{intro}
\begin{figure}[h]
\label{fig:simulations}
\includegraphics[width=0.9\linewidth]{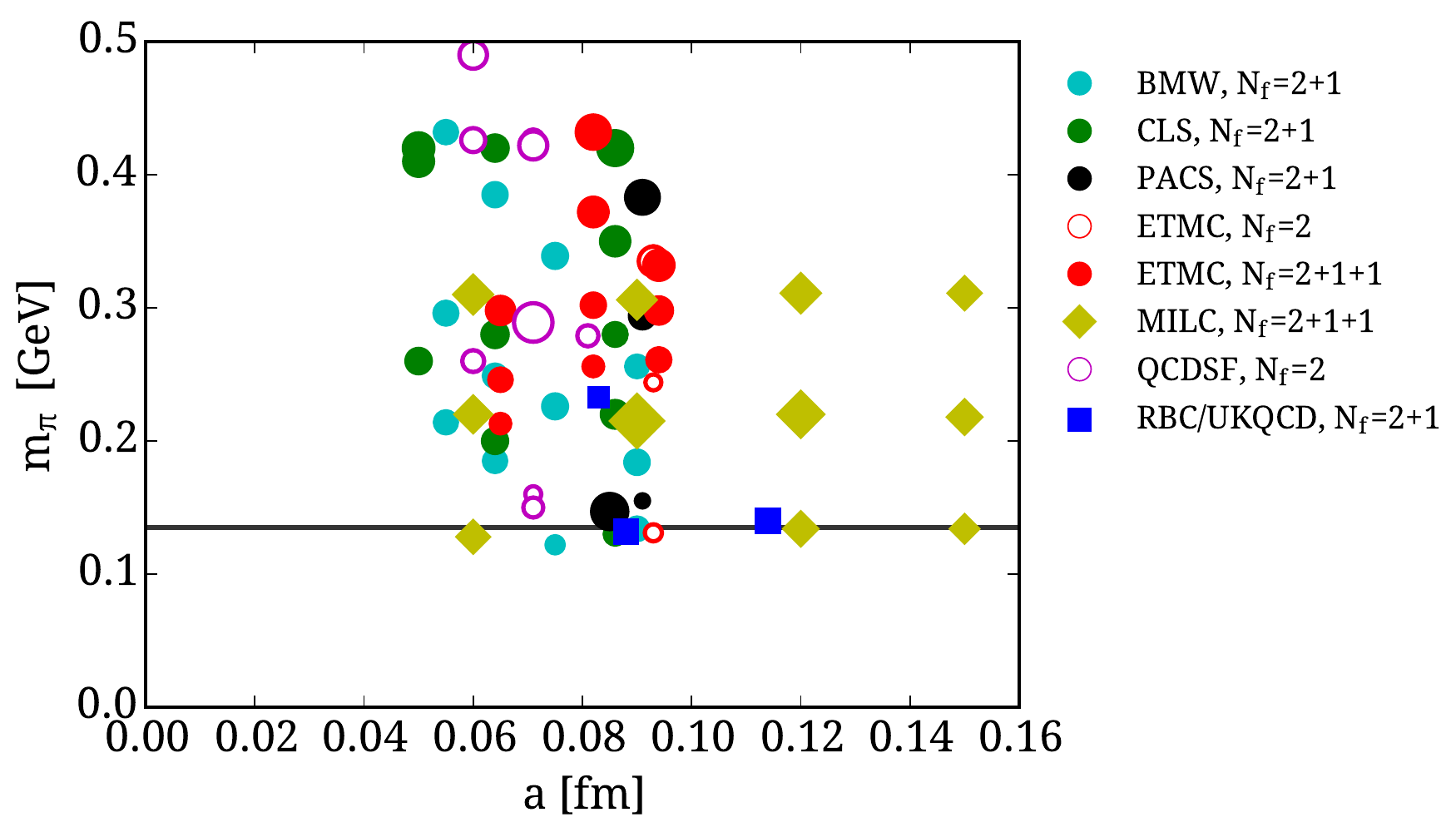} 
\caption{A summary of recent simulations showing the value of the pion mass and the lattice spacing: Black filled circles are from PACS using $N_f=2+1$ clover-improved fermions~\cite{Aoki:2009ix,Ishikawa:2015rho}, light blue filled circles from BMW for $N_f=2+1$ clover-improved fermions with HEX smearing~\cite{Durr:2010aw}, yellow filled diamonds from MILC using $N_f=2+1+1$ staggered fermions~\cite{Bazavov:2012xda}, magenta open circles from QCDSF using $N_f=2$ clover-improved fermions~\cite{Horsley:2013ayv}, green filled circles from CLS using $N_f=2+1$ clover-improved fermions~\cite{Bruno:2014jqa}, blue filled squares from RBC-UKQCD using domain wall fermions~\cite{Boyle:2015hfa}, red filled (open) circles from ETMC using $N_f=2+1+1$ twisted mass fermions ($N_f=2$ with a clover term)~\cite{Abdel-Rehim:2015pwa}. The size of the symbols is according to the value of $m_\pi L$ with the smallest value taken as $m_\pi L\sim 3$ and the largest $m_\pi L\sim 6.7$.}
\vspace*{-0.5cm}
\end{figure}

Lattice QCD simulations have seen tremendous progress in the last decade. This enables us to compute physical quantities more accurately but also to expand the
range of observables that can be extracted within the lattice QCD framework.
Simulations with quark masses fixed to their physical values, large enough volumes and small enough lattice spacings have become available  for a number of discretization schemes~\cite{Aoki:2009ix,Durr:2010aw,Bazavov:2012xda,Horsley:2013ayv,Bruno:2014jqa,Boyle:2015hfa,Abdel-Rehim:2015pwa}. In Fig.~\ref{fig:simulations} we show 
the status of recent simulations that include  various types of  Wilson ${\cal O}(a)$-improved, domain wall and staggered fermions~\cite{koutsou} simulated at a physical value of the pion mass and lattice spacing less that 0.1~fm where discretization errors for the light quark sector are found to be small.  Regarding finite volume effects
in the figure we include only simulations  with 
$m_\pi L \stackrel{>}{\sim} 3$ for which volume effects are expected to be under control for most of the quantities discussed here. 

\begin{figure}[h!]
\begin{minipage}{0.49\linewidth}
\includegraphics[width=\linewidth]{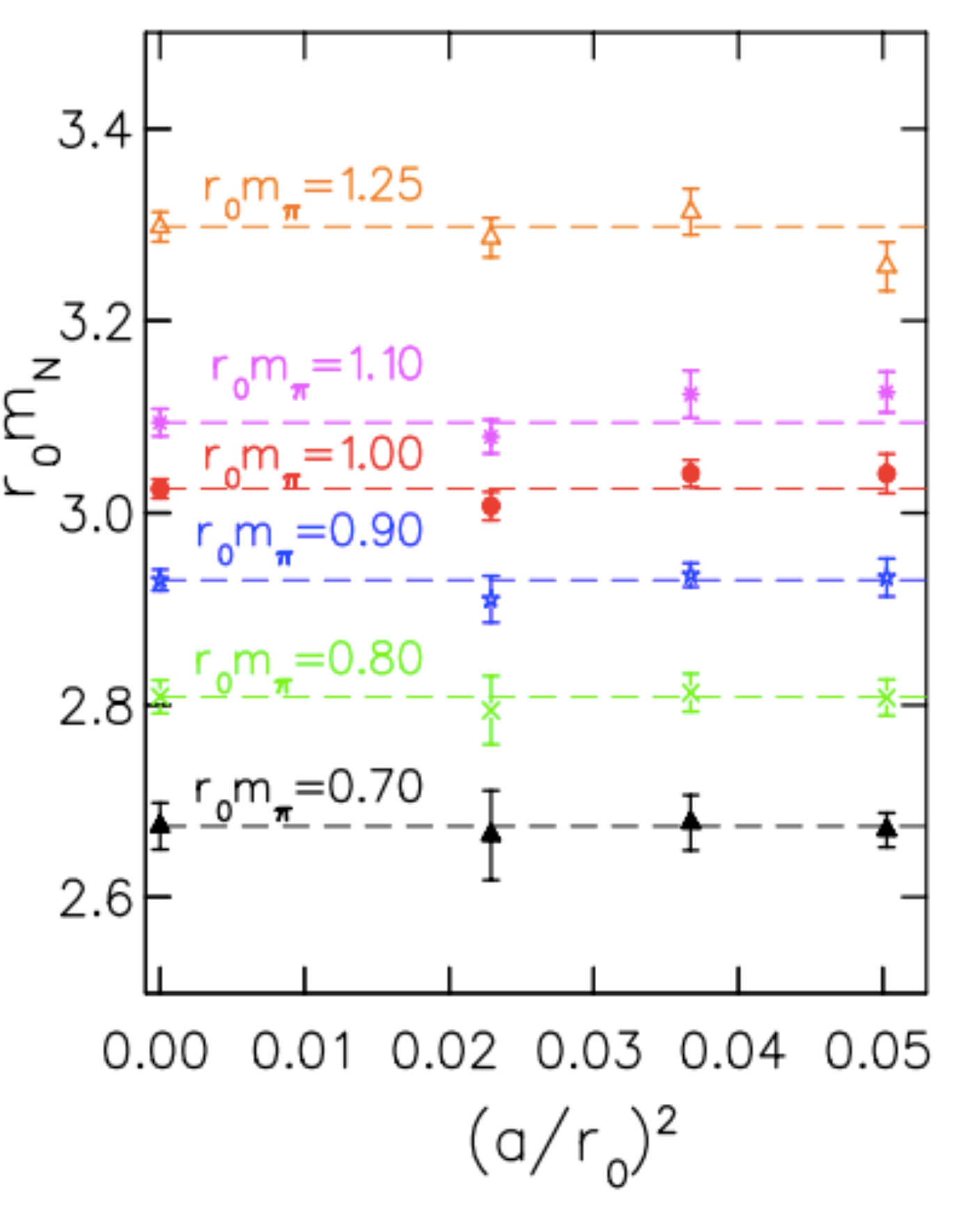}
\end{minipage}\hfill
\begin{minipage}{0.49\linewidth}
\includegraphics[width=\linewidth]{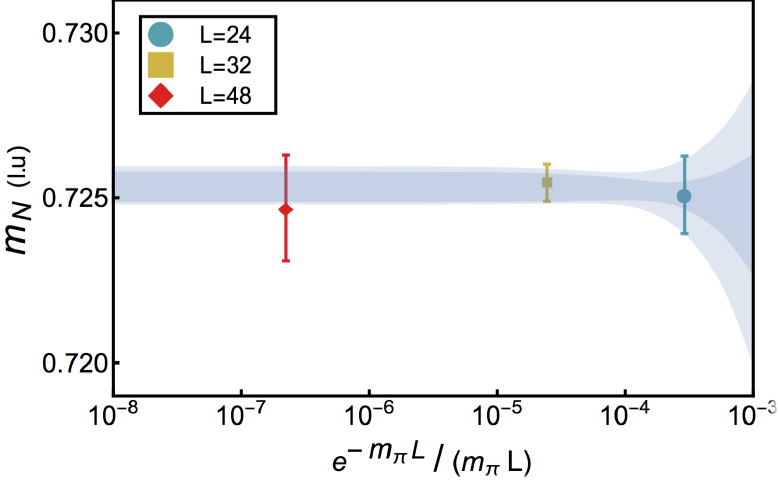}\\
{\includegraphics[width=\linewidth]{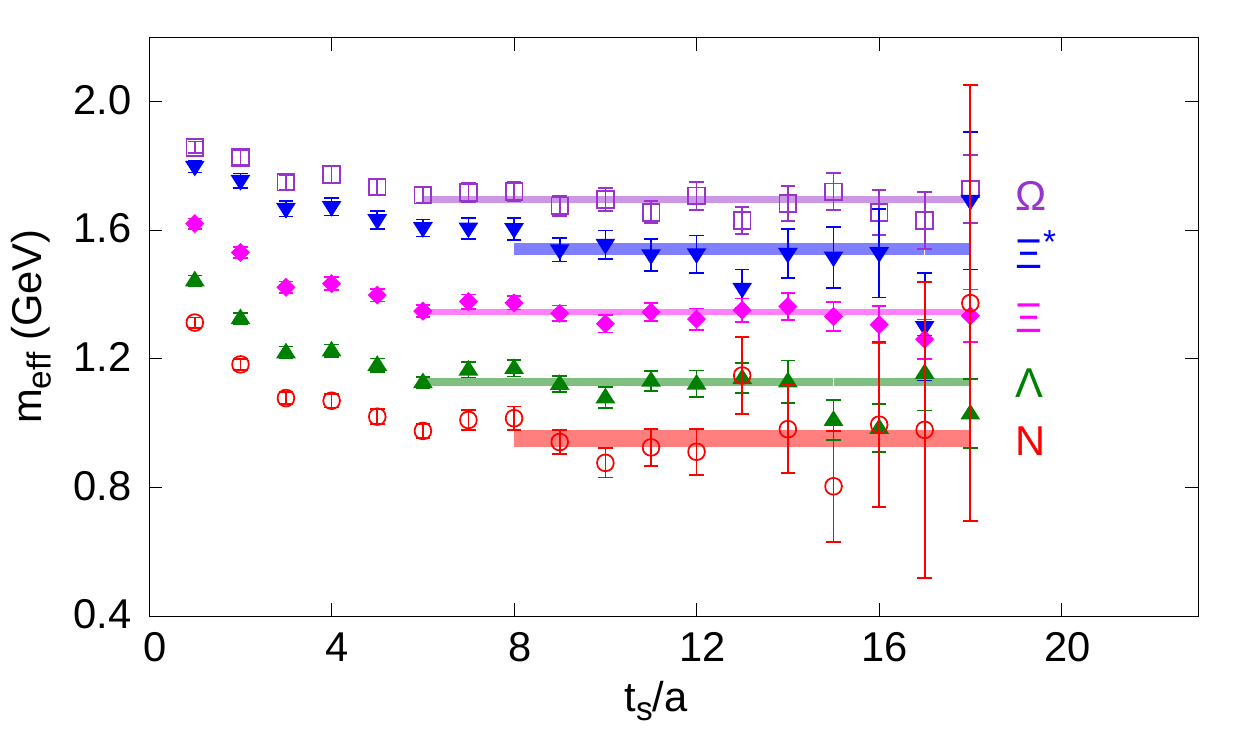}}
\end{minipage}
\caption{Left: The  nucleon mass versus $a^2$ for $N_f=2$  twisted mass fermions (TMF) for various pion masses in units of $r_0=0.44$~fm (ETMC)~\cite{Alexandrou:2008tn}. Right top:
Volume dependence of the nucleon mass for $m_\pi\sim 450$~MeV, $N_f=2+1$ Clover and $a\sim 0.12$~fm (NPLQCD)~\cite{Orginos:2015aya}. Right bottom:
Effective masses of baryons using $N_f=2$ twisted mass clover-improved fermions at physical pion mass~\cite{Alexandrou:2014wca}.}
\label{fig:a and V}
\vspace*{-0.5cm}
\end{figure}

For any lattice QCD analysis systematic uncertainties need to be carefully investigated in order
to compare with experimental values. These comprise of:
i) the finite lattice spacing $a$, where one needs at least three values in order to take the continuum limit $a\rightarrow 0 $; ii)
the finite spatial size of the lattice $L^3$  where at least three volumes are need  to estimate the infinite volume limit $L\rightarrow \infty$. Both these lattice artifacts have been typically
studied at larger than physical pion masses. For simulations using the physical value of the
light quark mass many results, especially beyond hadron masses, have been mostly
computed for one ensemble relying on assessing these systematics on the investigations done at larger pion masses. These so called {\it physical ensembles} eliminate the need for chiral extrapolations, which especially for the baryon sector, had in the past introduced systematic errors that typically dominated other systematic errors. In addition, 
iii) the identification of the hadron state 
of interest, which in particular for three-point correlators can introduce systematic errors due to higher states contributions that have to be carefully investigated. While cut-off and volume effects require the simulation of 
different gauge ensembles,  excited state contributions is done during the analysis on the same configurations.
We note that the inclusion of disconnected quark loop contributions has become feasible only recently eliminating an up to now uncontrolled approximation in hadron matrix elements.
In Fig.~\ref{fig:a and V} we show the nucleon mass computed for various
values of the lattice spacing  $a$  and lattice volume. As can be seen, for $a\stackrel{<}{\sim}0.1$~fm cut-off effects are small and for $L \stackrel{>}{\sim}3$ there are no visible volume effects at least for an ensemble with pion mass $m_\pi=450$~MeV.

\section{Hadron structure}\label{hadron structure}
The masses of the low-lying hadrons are well-studied with various discretization schemes using simulations summarized in Fig.~\ref{fig:simulations} where the continuum limit and finite volume effects have been investigated.
The mass is extracted via the Euclidean correlation function
\be
     G^H(\vec q, t_s) =\sum_{{\vec x}_s} \, e^{-i\vec {x}_s \cdot \vec q}\, 
     \langle J_H(\vec {x}_s,t_s)J_H^\dagger(0) \rangle  = \sum_{n=0,\cdots, \infty} A_{H_n} e^{-E_{H_n}(\vec{q})t_s}
\stackrel{t_s\rightarrow \infty} {\longrightarrow}A_{H_0} e^{-E_{H_0}(\vec{q}) t_s}\stackrel{\vec{q}=\vec{0}}{\longrightarrow}A_{H_0} e^{-m_{H_0} t_s},
\ee
where
the interpolating field has $J_H$ the quantum numbers of the hadron $H$ e.g. for $\pi^+$: $J_{\pi^+}(x)=\bar{d}(x)\gamma_5 u(x)$
and for the proton: $J_p(x)=\epsilon^{abc}\left(u^{a\top}(x)C\gamma_5 d^b(x)\right)u^c(x)$. 
The limit ${t_s\rightarrow \infty}$ in conjunction with the fact  than the noise to signal increases with $t_s \sim  e^{(m_H-\frac{3}{2}m_\pi)t_s}$ for baryons, means that an optimal time interval has to be identified for extracting the mass. 
Optimizing $J_H$ using smearing techniques to maximize its overlap with to the lowest state is essential in order to achieve early convergence. 
Defining  
\be    a E_{\rm eff}(\vec q, t_s) \equiv \ln \left[ \frac{G^H(\vec q, t_s)}{G^H(\vec q, t_s+a)}\right]
= aE_{H_0}(\vec{q})+ {\rm excited \>states}\stackrel{t_s\rightarrow \infty} {\longrightarrow} aE_{H_0}\stackrel{{\vec q}=0}{\to} am_{H_0}
\label{eff mass}
\ee
we can extract the mass of the hadron $H$. Representative examples of the behavior of baryon effective masses using simulations with a physical value of the
pion mass is shown in Fig.~\ref{fig:a and V}.

The evaluation of hadron matrix elements requires the computation of the
appropriate  Euclidean three-point function,
$G^{\mu\nu}(\Gamma,\vec q,t_s, t_{\rm ins}) =\sum_{\vec x_s, {\vec x}_{\rm ins}} \, e^{i{\vec x}_{\rm ins} \cdot \vec q}\, 
     {\Gamma_{\beta\alpha}}\, \langle {J^{\alpha}_H(\vec x_s,t_s)} {\cal O}_\Gamma^{\mu\nu}({\vec x}_{\rm ins},t_{\rm ins}) {\overline{J}^{\beta}_H(\vec{x}_0, t_0)} \rangle $,
and  dividing it by an appropriate combination of two-point functions such that, at large Euclidean times, the ratio yields the matrix element of interest:
 \begin{align}
    R(t_s,t_{\rm ins},t_0) \xrightarrow[(t_s-t_{\rm ins})\Delta \gg 1]{(t_{\rm ins}-t_0)\Delta \gg 1} \mathcal{M}[1
      +  \mathcal{T}_1e^{-\Delta({\bf p})(t_{\rm ins}-t_0)} + \mathcal{T}_2 e^{-\Delta({\bf
          p'})(t_s-t_{\rm ins})}+\cdots],
  \end{align}
where 
 $\mathcal{M}$ the desired matrix element,  $t_s,t_{\rm ins},t_0$ is the
  sink, insertion and source times and $\Delta({\bf p})$ the
  energy gap with the first excited state.

\subsection{Nucleon scalar, axial and tensor charges}
\begin{figure}[h!]
\begin{minipage}{0.49\linewidth}
\includegraphics[width=\linewidth]{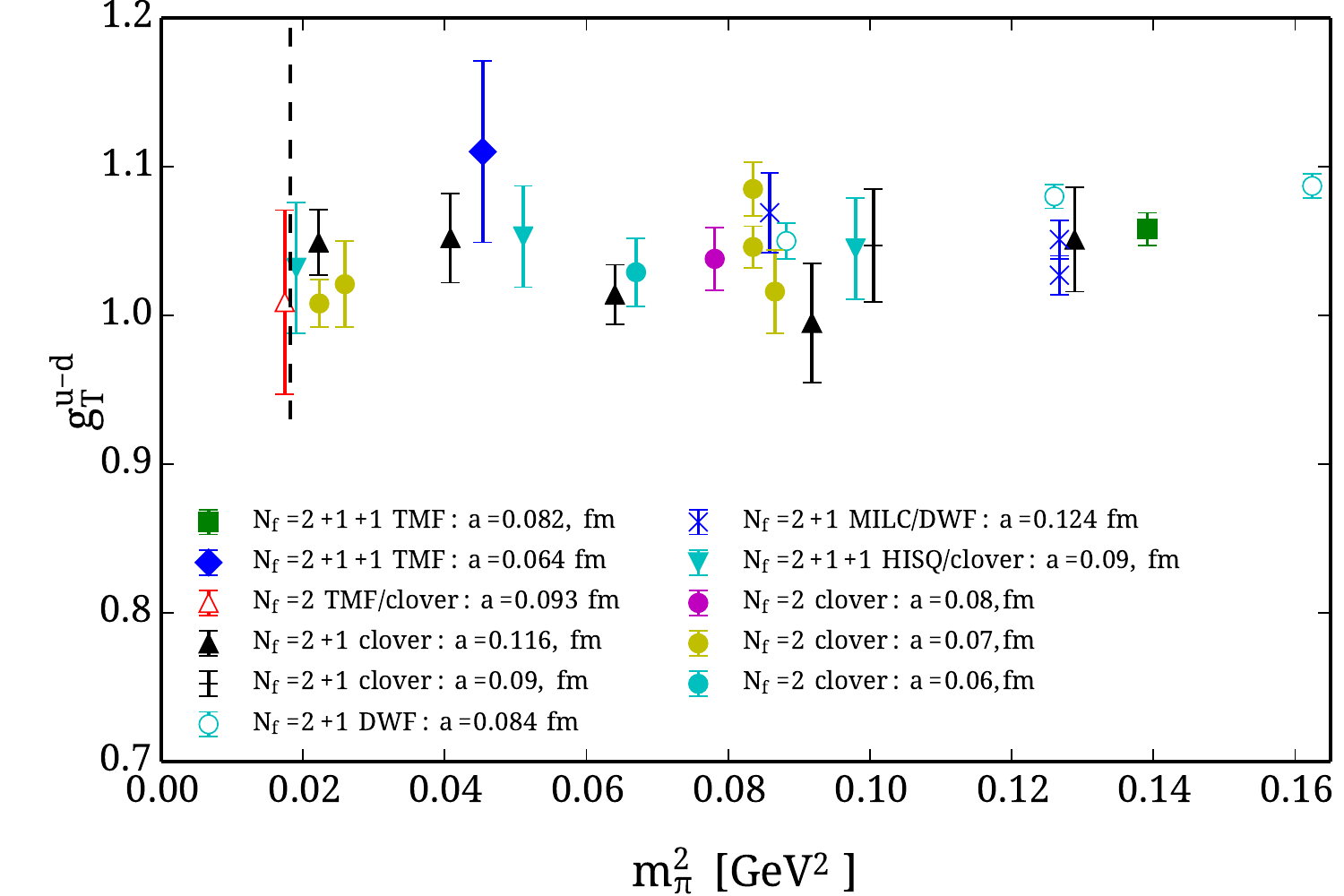}
\end{minipage}\hfill
\begin{minipage}{0.49\linewidth}
\includegraphics[width=\linewidth]{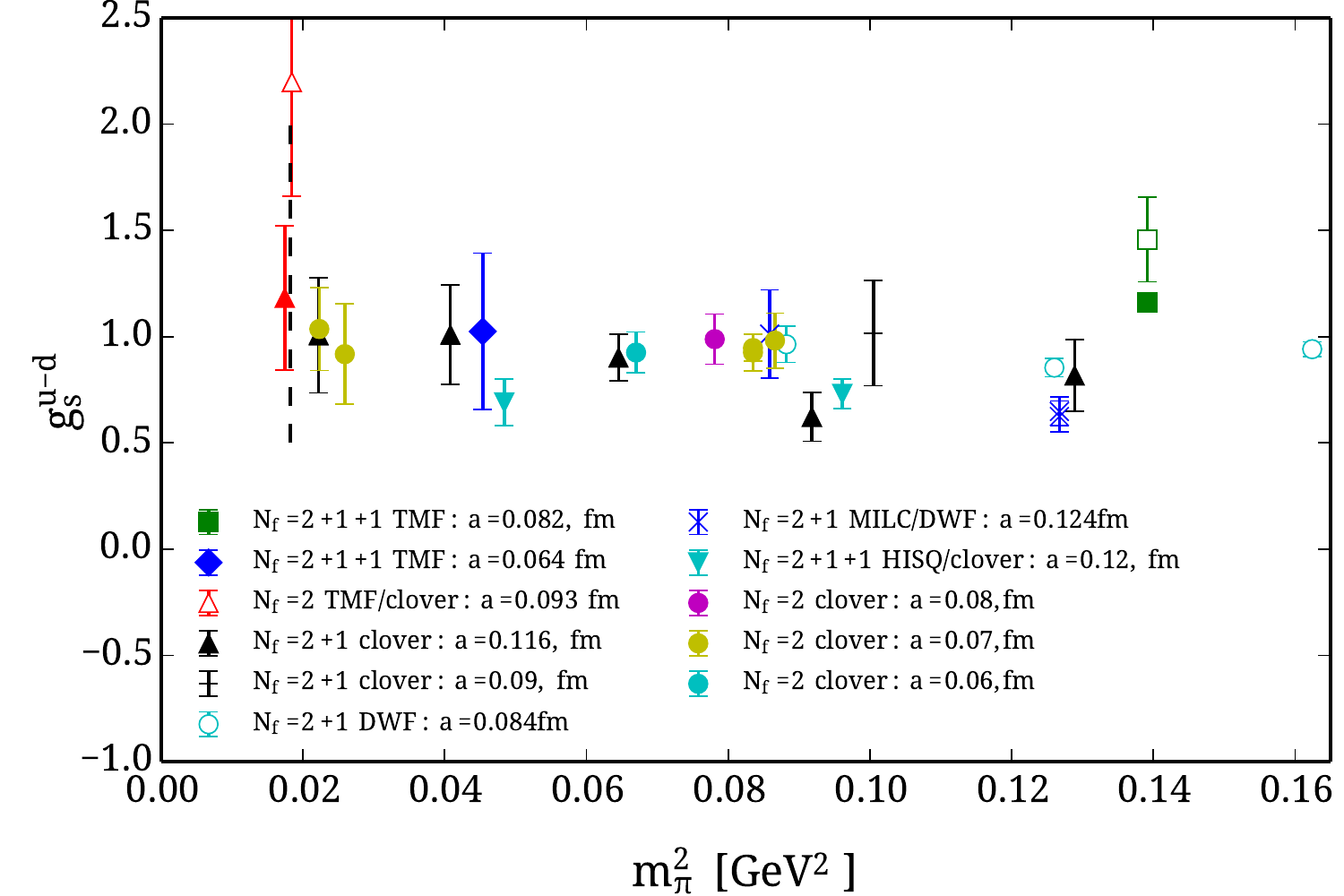}
\end{minipage}
\caption{Left panel: The nucleon isovector tensor charge. Right panel: The isovector scalar charge.}
\label{fig:gT and gs}
\vspace*{-0.5cm}
\end{figure}

The nucleon matrix element of the axial-vector operator ${\cal O}_{A}^3=\bar{\psi}(x)\gamma^{\mu}\gamma_5\frac{\tau^3}{2}\psi(x)$ at zero momentum transfer 
yields the well-known nucleon axial charge $g_A$, measured in neutron $\beta$-decay. 
Because of the isovector nature
of the axial-vector only connected contributions are non-vanishing in the isospin limit.  
\begin{figure}[h!]
\begin{minipage}{0.49\linewidth}
\includegraphics[width=\linewidth]{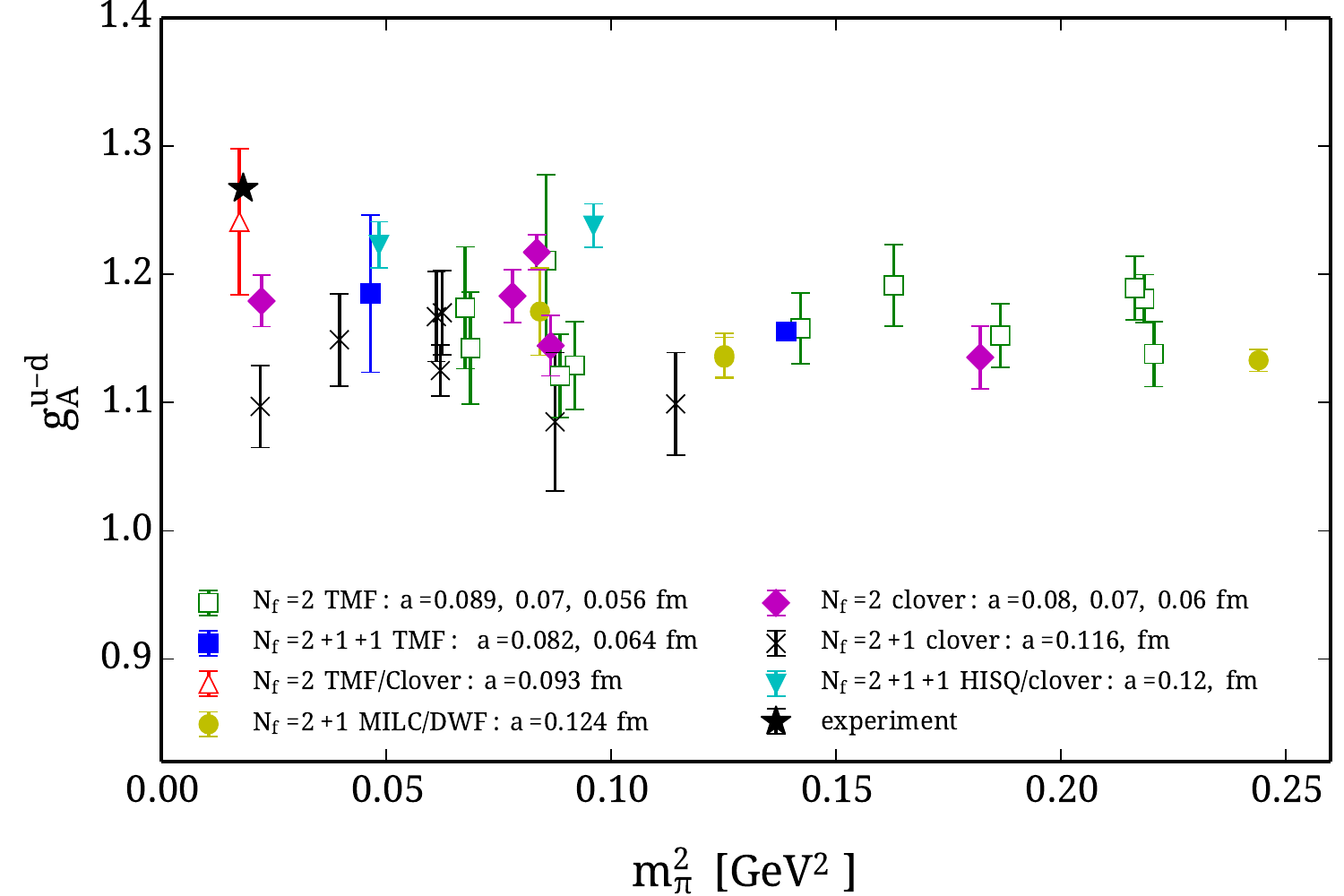}
\caption{Lattice QCD results on the nucleon $g_A$.}
\label{fig:gA}
\end{minipage}\hfill
\begin{minipage}{0.49\linewidth}\vspace*{-1.3cm}
Due to the relative ease to compute $g_A$ and  its accurately measured value  it has been studied extensively in lattice QCD as a benchmark quantity. In Fig.~\ref{fig:gA} we collect results from various collaborations. As can be seen, the value obtained using simulations with a physical value of the pion mass agrees with the experimental one, while for larger pion masses it was
underestimated by all groups. This clearly demonstrates the importance of
these simulations.
\end{minipage}
\vspace*{-0.5cm}
\end{figure}
Having reproduced $g_A$, the less known isovector tensor and   scalar charges can be evaluated using similar techniques. The corresponding currents are  ${\cal O}_{T}^3=\bar{\psi}(x)\sigma^{\mu\nu}\frac{\tau^3}{2}\psi(x)$ for the tensor and
${\cal O}_{S}^3=\bar{\psi}(x)\frac{\tau^3}{2}\psi(x)$ for the scalar operators.  In Fig.~\ref{fig:gT and gs} we collect results from various collaborations.
We note that the experimental value of $g_T^{u-d}\sim 0.54 ^{+0.30}_{-0.13}$ resulting from a global analysis of HERMES, COMPASS and Belle $e^+ e^-$ data~\cite{Anselmino:2013vqa}, while a new analysis of COMPASS and Belle data yields $g_T^{u-d}=0.81(44)$~\cite{Radici:2015mwa}. Given this large uncertainty, a lattice QCD determination can provide valuable input, especially in view of plans to measure $g_T$ in the  SIDIS experiment  on $^3$He/Proton  at 11 GeV at JLab.
The scalar charge shows large excited states contributions  
 and a larger $t_s-t_0$ is required as compared to e.g. $g_T$ in order to extract the correct matrix element.
We found that  $t_s-t_0\stackrel{>}{\sim} 1.5$~fm is needed for convergence.

\subsection{Axial charges of other baryons}
Besides the axial charge for the nucleon, axial charges of other particles can also be computed. Many of these are
difficult or even not feasible to measure experimentally and lattice QCD can provide valuable information on these couplings, which enter in chiral Lagrangians.

\begin{figure}[h!]
\begin{minipage}{0.49\linewidth}
\includegraphics[width=\linewidth]{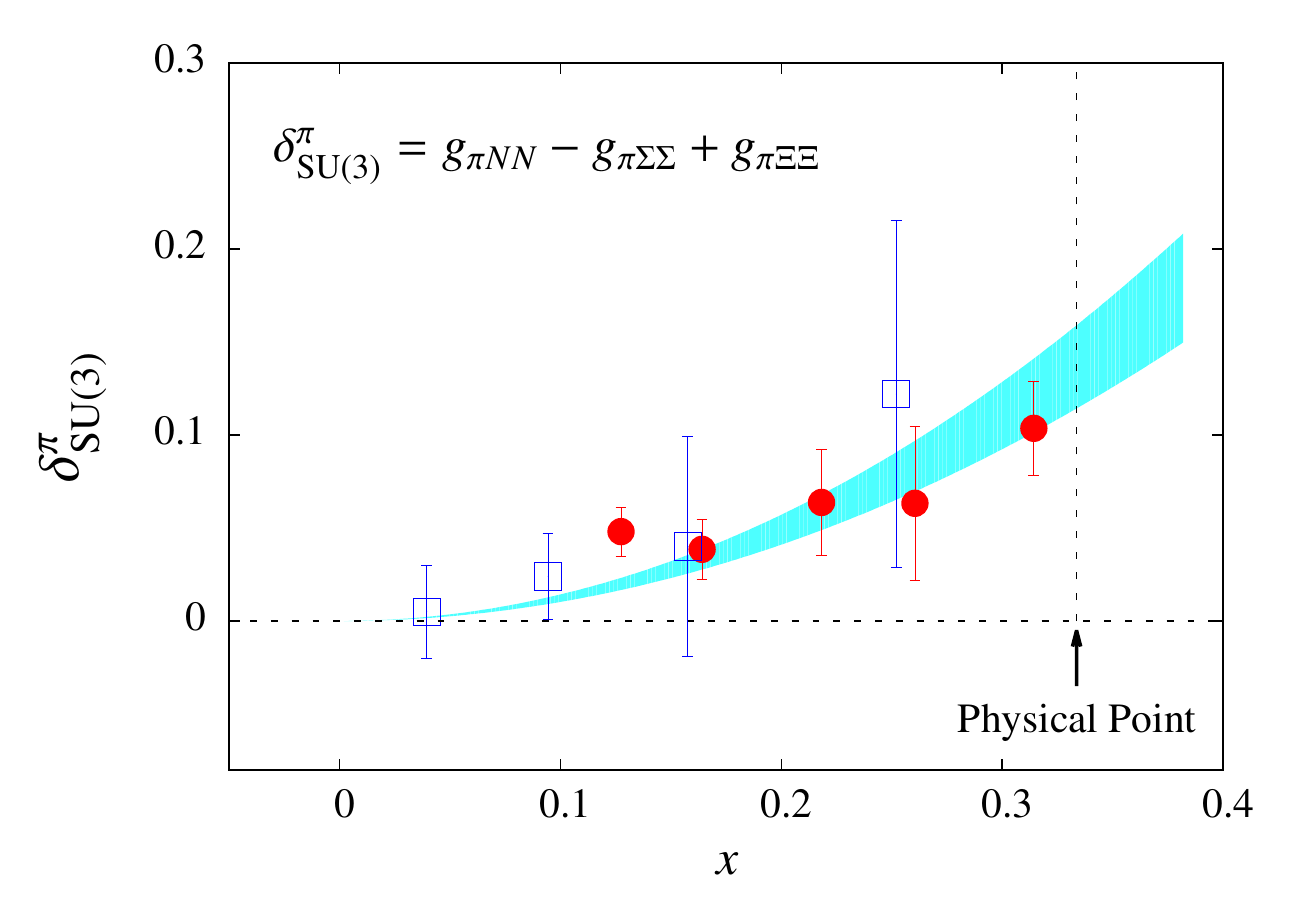}
\end{minipage}\hfill
\begin{minipage}{0.47\linewidth}
\includegraphics[width=\linewidth]{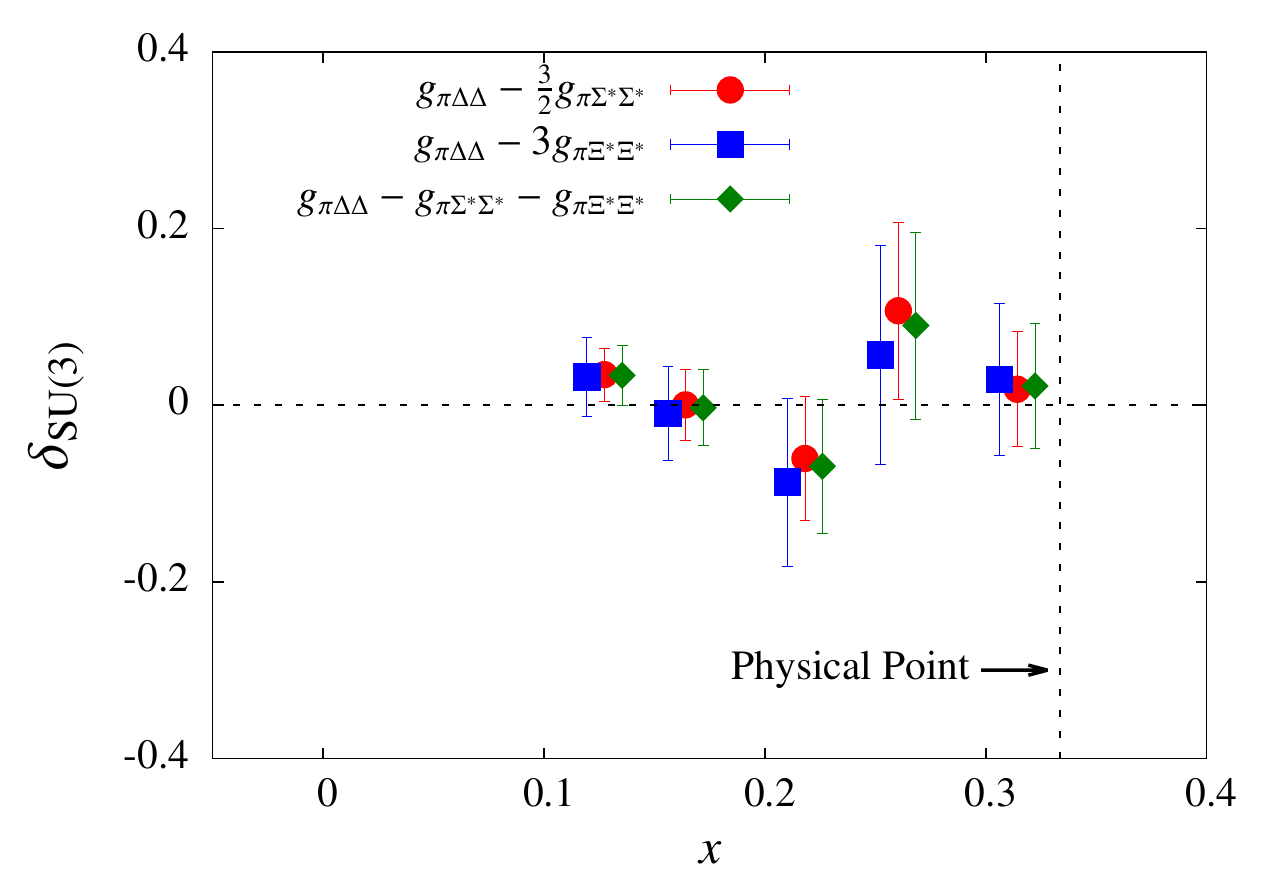}
\end{minipage}
\caption{The SU(3) breaking parameter $\delta_{\rm SU(3)}$ for the octet (left) and the decuplet (right). Results are from Refs.~\cite{Lin:2007ap,Alexandrou:2014wca}.}
\label{fig:hyperon axial}
\vspace*{-0.5cm}
\end{figure}
In Fig.~\ref{fig:hyperon axial} we show the SU(3) breaking parameter $\delta_{\rm SU(3)}=g_A^N-g_A^\Sigma +g_A^\Xi$ versus $x=(m_K^2-m_\pi^2)/4\pi^2 f_\pi^2$ for the octet baryons. As can be seen, the SU(3) breaking is about 10\%-15\% at
the physical point. The axial charges for the decuplet baryons are given in terms of one coupling constant in the SU(3) limit leading to three relations. These
relations are shown in the same figure and show no detectable SU(3) breaking to the accuracy of our data.

\subsection{Generalized Parton Distributions }
Another set of observables that probes the structure of hadrons are Generalized Parton Distributions (GDPs) measured in deep inelastic scattering.
These are matrix elements in the infinite momentum frame but
factorization leads to a set of three twist-two local operators, namely the vector operator 
${\cal O}_{V^a}^{\mu_1 \cdots \mu_n}=\bar{\psi}(x)\gamma^{\{\mu_1}i\Dlr^{\mu_2}\ldots i\Dlr^{\mu_n \}}\frac{\tau^a}{2}\psi(x)$, the  axial-vector operator
${\cal O}_{A^a}^{\mu_1 \cdots \mu_n}=\bar{\psi}(x)\gamma^{\{\mu_1}i\Dlr^{\mu_2}\ldots i\Dlr^{\mu_n \}}\gamma_5\frac{\tau^a}{2}\psi(x)$ and the  tensor operator
${\cal O}_{T^a}^{\mu_1 \cdots \mu_n}=\bar{\psi}(x)\sigma^{\{\mu_1,\mu_2}i\Dlr^{\mu_3}\ldots i\Dlr^{\mu_n \}}\frac{\tau^a}{2}\psi(x)$.
In the special case where  we have  no derivatives these yield the usual hadron form factors, while for zero momentum transfered squared $q^2=-Q^2=0$ they reduce to the  parton distribution functions (PDFs) yielding for  instance 
the  average momentum fraction or unpolarized moment $\langle x \rangle$  in the case of  the one-derivative vector operator.

For a spin-1/2 particle, like the nucleon, the decomposition of the matrix element of the one-derivative vector operator is given by

\be 
\langle N(p^\prime,s^\prime) | {\cal O}_{V^3}^{\mu\nu}| N(p,s) \rangle = 
    \bar u_N(p^\prime, s^\prime) 
     \Biggl[  {A_{20}(q^2)} \gamma^{\{\mu}P^{\nu\}}+{B_{20}(q^2)} \frac{i\sigma^{\{\mu \alpha}q_\alpha P^{\nu\}}}{2m}
+C_{20}(q^2) \frac{q^{\{\mu}q^{\nu\}}}{m} \biggr] \frac{1}{2}u_N(p,s) \quad.
\label{vector derivative}
\ee
Extracting $A_{20}$ and $B_{20}$ is particularly relevant for  understanding the nucleon spin $J^q$ carried by a quark since $J^q=\frac{1}{2}\biggl [A^q_{20}(0)+B^q_{20}(0)\biggr]$ as well as the momentum fraction  $\langle x \rangle_q=A^q_{20}(0)$.

\begin{figure}[h!]
\begin{minipage}{0.49\linewidth}
\includegraphics[width=\linewidth]{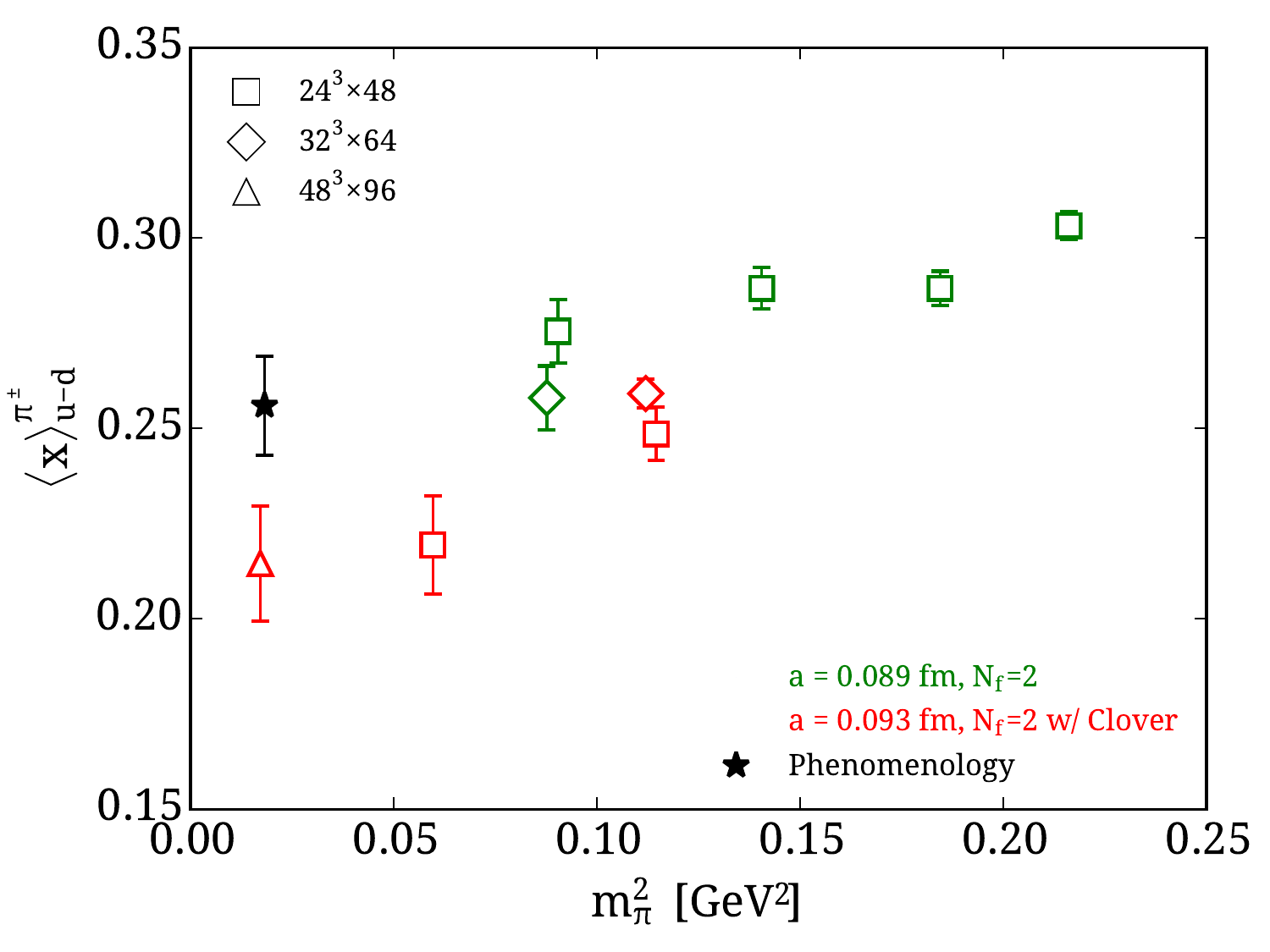}
\end{minipage}
\begin{minipage}{0.49\linewidth}
\includegraphics[width=\linewidth]{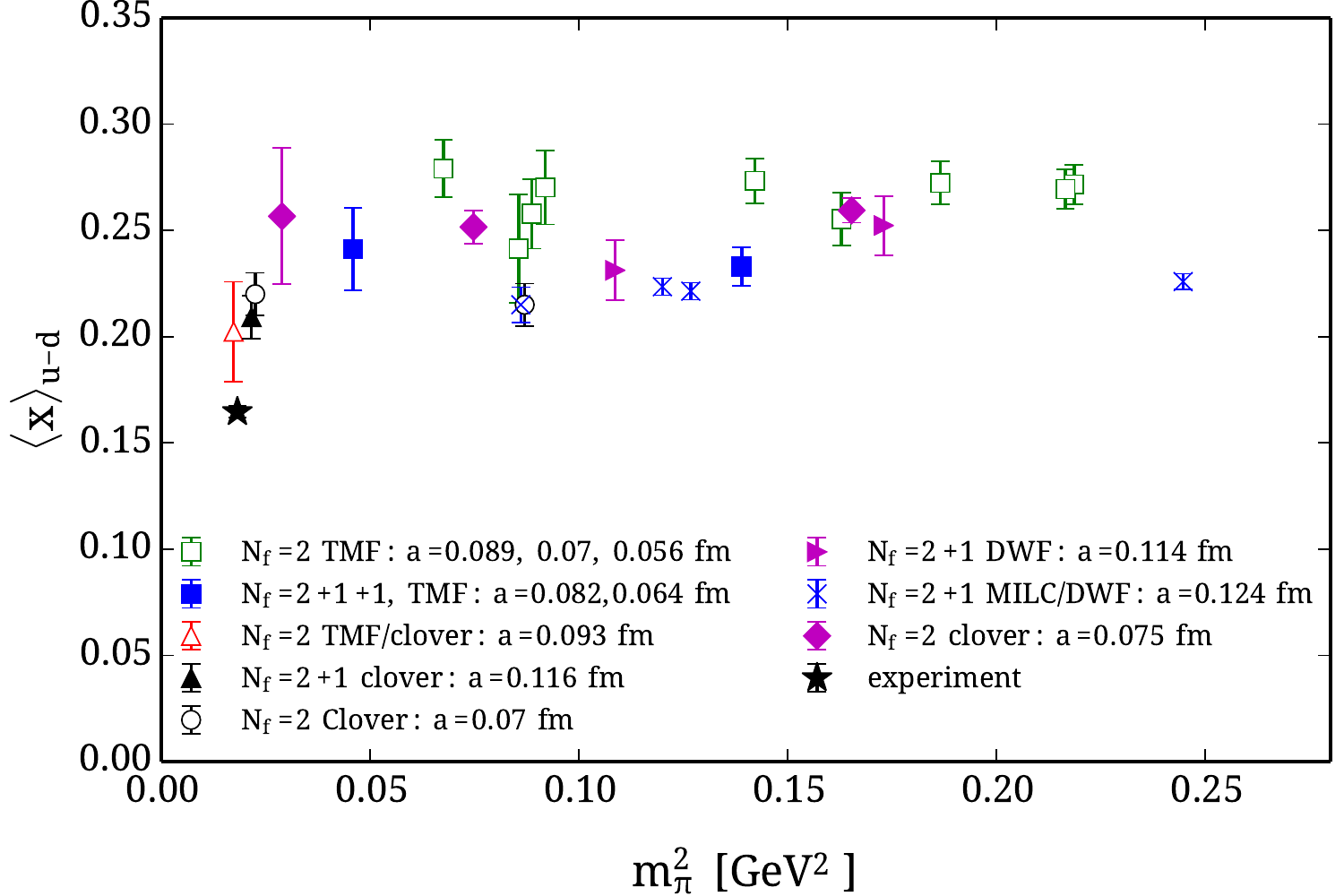}
\end{minipage}
\caption{The isovector momentum fraction $\langle x\rangle_{u-d}$  for the pion (left) and the nucleon (right).
The lattice QCD results are from Refs.~\cite{Abdel-Rehim:2015pwa,Green:2012ud}. The experimental values are from Ref.~\cite{Wijesooriya:2005ir} and \cite{Alekhin:2012ig}, respectively.}
\label{fig:ave. x}
\vspace*{-0.5cm}
\end{figure}

\noindent
{\it Momentum fraction:} In Fig.~\ref{fig:ave. x} we show $\langle x\rangle_{u-d}$  obtained in the $\overline{MS}$ scheme at  $\mu= 2$~GeV for the pion and the nucleon. While volume effects are not statistically significant at larger than physical pion mass where we have more than one lattice volume, at the physical   finite volume effects  have not been investigated yet and they could be the reason  for the small discrepancy with the experimental value especially for the pion. For the nucleon, excited state contributions could be another reason, which is being investigated.

\noindent
{\it Nucleon gluon unpolarized moment:} We have also computed
the matrix element  $\langle N|O_{44}-\frac{1}{3}O_{jj}|N\rangle$ at zero momentum, which yields directly $\langle x \rangle_g$, where we considered
the gluon operator $O_{\mu\nu}=-{\rm Tr}[ G_{\mu\rho} G_{\nu\rho}]$.
We used  HYP-smearing to reduce noise and  perturbative renormalization.
The analysis was carried out using an ensemble of $N_f=2+1+1$ TMF with $a$ = 0.082~fm, $m_\pi$ = 373~MeV and $\sim$ 34,470 statistics~\cite{Alexandrou:2013tfa} as well as with an ensemble of
 $N_f=2$ TMF plus clover, $a=0.093$~fm, $m_\pi=132$~MeV and $\sim$155,800 statistics.
We find $\langle x\rangle_g=0.282(39)$ for the physical ensemble in ${\overline{\rm MS}}$ at $\mu=2$~GeV.

\begin{figure}[h!]
\begin{minipage}{0.49\linewidth}
{\includegraphics[width=\linewidth]{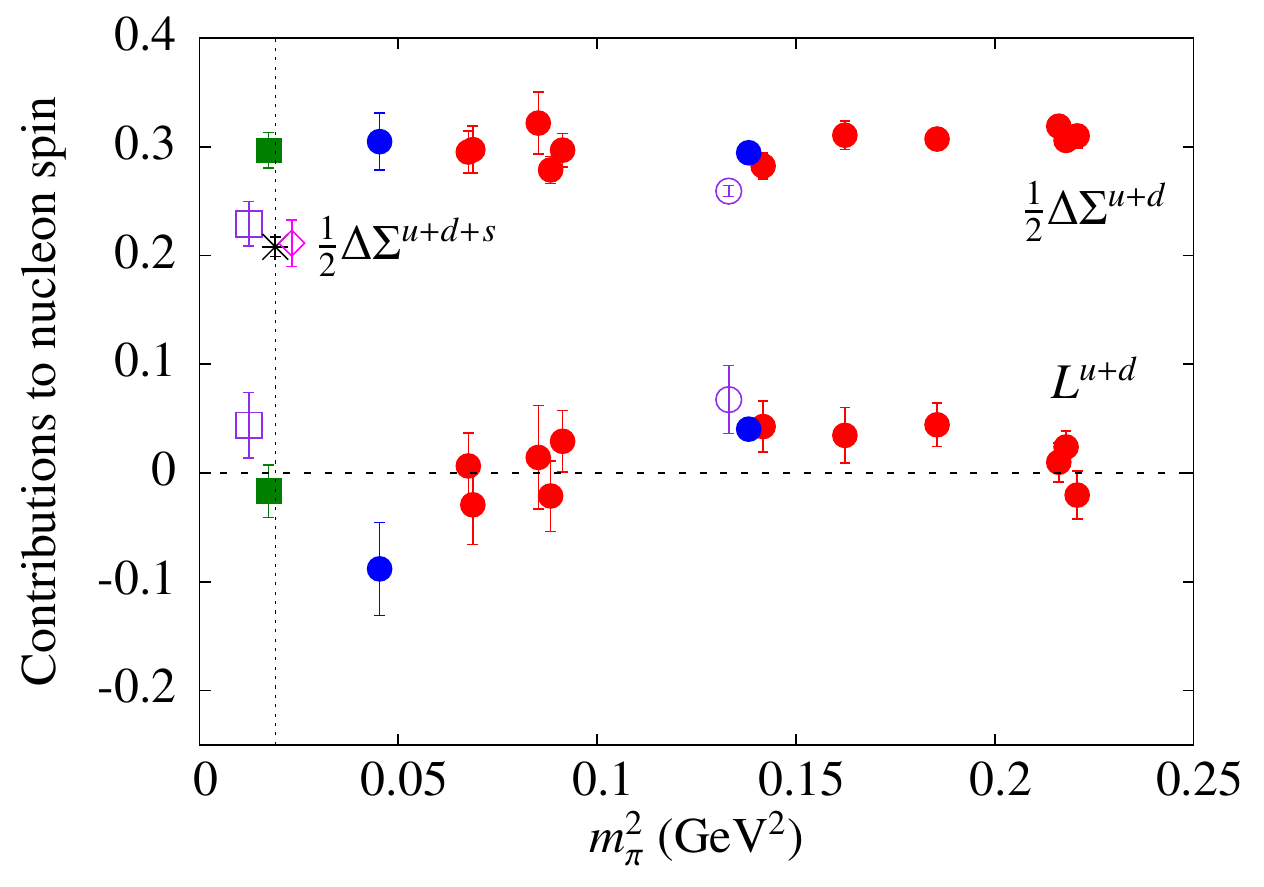}}
\end{minipage}
\begin{minipage}{0.49\linewidth}
{\includegraphics[width=\linewidth]{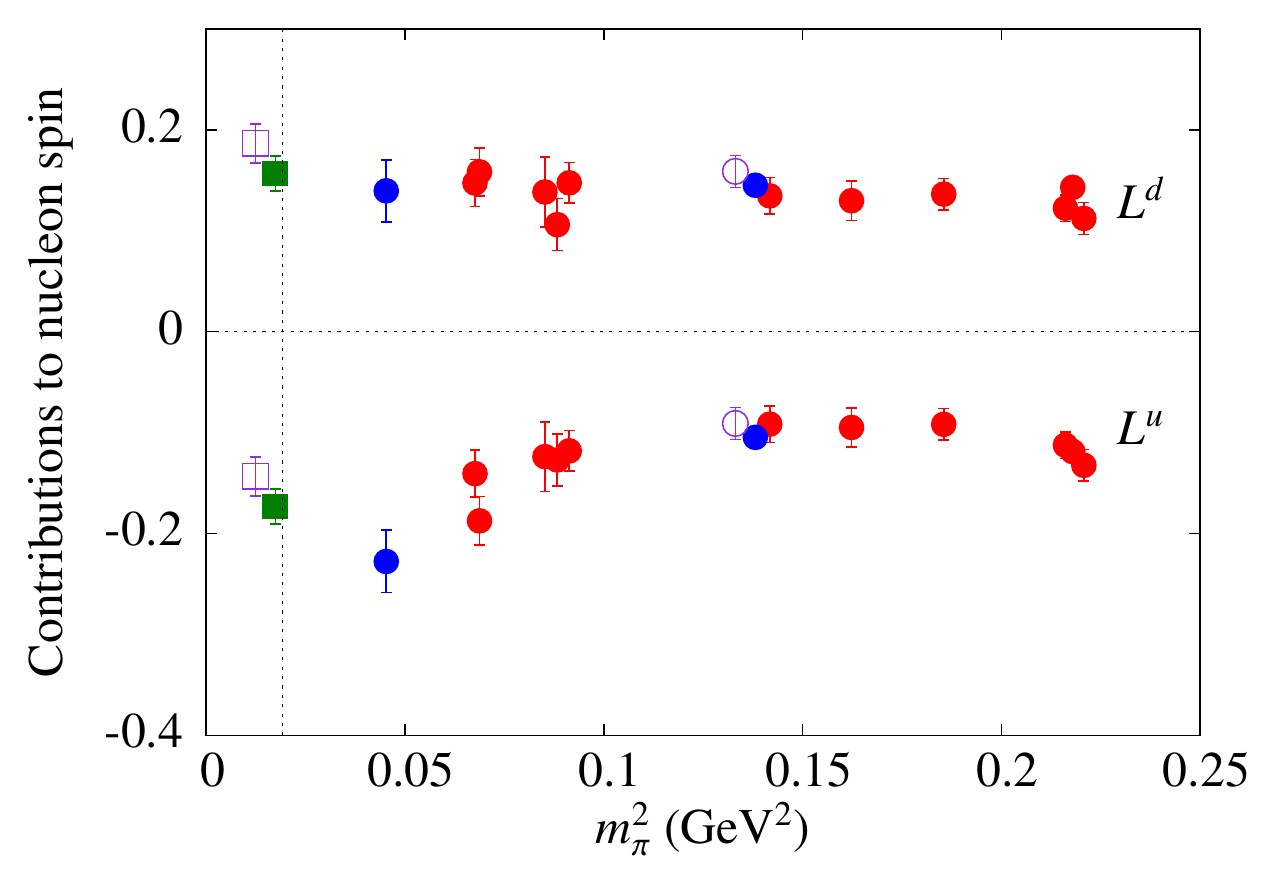}}
\end{minipage}
\caption{$\Delta \Sigma^{u+d}$ (left) and $L^{u,d}$ (right) in the $\overline{\rm MS}$ at 2~GeV using $N_f=2$ and $N_f=2+1+1$ twisted mass fermions. Open squares  include disconnected contributions from the u and d quarks, while the open diamond also includes the strange quark contribution to  $\Delta \Sigma$.}
\label{fig:nucleon spin}
\vspace*{-0.5cm}
\end{figure}

\noindent
{\it Nucleon spin:} The nucleon spin can be written as $ \frac{1}{2}=\sum_{q}J^q={\left(\frac{1}{2}\Delta \Sigma^q +L^q\right)} +J^G $, where
$\Delta \Sigma^q=g_A^q$. 
Disconnected contributions have been computed using  ${\cal O} (150, 000)$ statistics for an ensemble of $N_f=2+1+1$ TMF at $m_\pi=373$~MeV~\cite{Abdel-Rehim:2013wlz}, and for $N_f=2$ TMF with a clover term at  $m_\pi=132$~MeV~\cite{Abdel-Rehim:2015lha}. In Fig.~\ref{fig:nucleon spin} we show TMF results for $\Delta \Sigma^q$ and $L^q$.  At the physical point, we find a  value of  $J^{u+d}= 0.273(22)$ and $L^u\sim -L^d$, while  $\frac{1}{2}\Delta \Sigma^{u+d}=0.229(20)$ and $\frac{1}{2}\Delta \Sigma^{u+d+s}=0.211(21)$, where  for the first time, disconnected contributions are included at the physical point bringing agreement with the experimental value.

\begin{figure}[h!]
\begin{minipage}{0.45\linewidth}
{\includegraphics[width=\linewidth,height=0.9\linewidth]{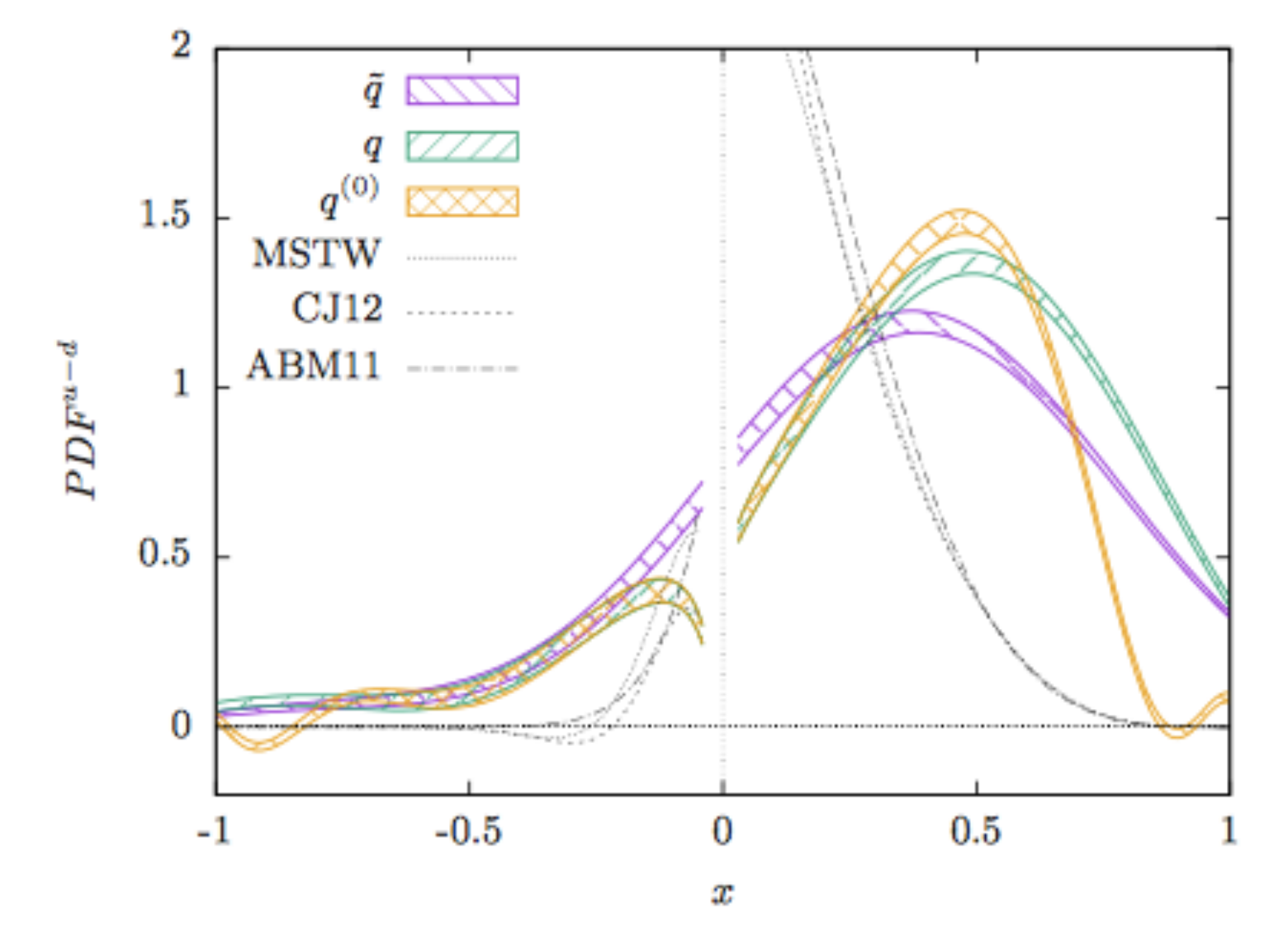}}
\caption{Results on the unrenormalized $q(x)$ for 5-HYP steps, $P_3=4\pi/L$ from Ref.~\cite{Alexandrou:2015rja}.}
\label{fig:PDF}
\end{minipage}\hfill
\begin{minipage}{0.54\linewidth}\vspace*{-1.3cm}
We note that: i)
The calculation of the leading UV divergences in $\tilde{q}$ perturbatively   is done keeping $P_3$ fixed while taking $\Lambda \to \infty$ (in contrast to first taking  $P_3\to\infty$ for the renormalization of $q$); ii)  The renormalization procedure is still under study and thus here 
we  identify the UV regulator as $\mu$ for $q(x)$
and as  $\Lambda$ for the case of the quasi-distribution $\tilde{q}(x)$. The dependence on the UV regulator $\Lambda$ will be translated, in the end, into a renormalization scale $\mu$ after proper renormalization; 
iii) Single pole terms cancel when combining the vertex and wave function corrections, and double poles are  reduced to a single pole that are taken care via the  principal value prescription;
iv) A divergent term remains in $\delta Z^{(1)}$ that depends on the cut-off $x_c$
\end{minipage}
\end{figure}

\noindent
{\it Direct evaluation of parton distribution functions - an exploratory study:}
We consider the matrix element: 
$\tilde{q}(x,\Lambda,P_3)=\int_{-\infty}^{+\infty}  \frac{dz}{4\pi} e^{-izxP_3}{\langle P|\bar{\psi}(z,0)\>\gamma_3 \,W(z)\psi(0,0)|P\rangle}_{h(P_3,z)} $ where $\tilde{q}(x)$
is the quasi-distribution defined in Ref.~\cite{Ji:2013dva}, which
 can be computed in lattice QCD. First results are obtained for $N_f=2+1+1$ clover fermions on HISQ sea~\cite{Lin:2014zya} and for an $N_f=2+1+1$ TMF ensemble with $m_\pi=373~$MeV~\cite{Alexandrou:2015rja} for which we show results in  Fig.~\ref{fig:PDF} on the isovector distribution 
$q^{u-d}(x)$ for 5 steps of HYP smearing. 
The matching to the PDF $q(x)$ is done using 
\be
q(x,\mu)=\tilde{q}(x,\Lambda,P_3)-\frac{\alpha_s}{2\pi}\tilde{q}(x, \Lambda,P_3)\delta Z_F^{(1)}\left(\frac{\mu}{P_3},\frac{\Lambda}{P_3}\right)-\frac{\alpha_s}{2\pi}\int_{-1}^1  \frac{dy}{y} Z^{(1)}\left(\frac{x}{y},\frac{\mu}{P_3},\frac{\Lambda}{P_3}\right)\tilde{q}(y,\Lambda,P_3) +{\cal O}(\alpha_s^2)
\label{PDF}
\ee

\vspace*{-0.5cm}

\section {Conclusions}
 Simulations at near physical parameters of QCD are beginning to yield important results on benchmark quantities such as the mass of the low-lying hadrons, the  nucleon axial charge and the pion decay constant. This well-established framework can thus be employed for predicting other quantities probing hadron structure such as   
 scalar and tensor charges,  tensor moments, and $\sigma$-terms. Exploration of new techniques to  compute hadron PDFs, charge radii and electric dipole moments is on-going,  as well as, 
the development of techniques for resonances 
and for {\it ab Initio} Nuclear Physics~\cite{Savage:2015eya}. This thus represents a very rich program for zero-temperature hadron and nuclear physics and we expect rapid progress in many of these areas in the near future.

\section*{Acknowledgments}
I would like to thank my collaborators A. Abdel-Rehim, M. Constantinou, K. Jansen, K. Hadjiyiannakou, Ch. Kallidonis, G. Koutsou, F. Steffens,  C. Wiese and A. Vaquero for their invaluable contributions.
This work was supported by a grant from the Swiss National Supercomputing Centre (CSCS) under project ID s540 and in addition used computational resources  from
the John von Neumann-Institute for Computing on the Juropa system and
the BlueGene/Q system Juqueen  partly through PRACE allocation, which included Curie (CEA), Fermi (CINECA) and SuperMUC (LRZ).


\begin{thebibliography}{}
%
%


\bibitem{Aoki:2009ix} 
  S.~Aoki {\it et al.} [PACS-CS Collaboration],
  Phys.\ Rev.\ D {\bf 81}, 074503 (2010)
    [arXiv:0911.2561].
\bibitem{Durr:2010aw} 
  S.~Durr {\it et al.},
  JHEP {\bf 1108}, 148 (2011)
    [arXiv:1011.2711].
\bibitem{Bazavov:2012xda} 
  A.~Bazavov {\it et al.} [MILC Collaboration],
  Phys.\ Rev.\ D {\bf 87}, no. 5, 054505 (2013)
    [arXiv:1212.4768].
\bibitem{Horsley:2013ayv} 
  R.~Horsley, Y.~Nakamura, A.~Nobile, P.~E.~L.~Rakow, G.~Schierholz and J.~M.~Zanotti,
  Phys.\ Lett.\ B {\bf 732}, 41 (2014)
    [arXiv:1302.2233];
  G.~S.~Bali {\it et al.},
  Phys.\ Rev.\ D {\bf 90}, no. 7, 074510 (2014)
  [arXiv:1408.6850].
\bibitem{Boyle:2015hfa} 
  P.~A.~Boyle {\it et al.} [RBC/UKQCD Collaboration],
  JHEP {\bf 1506}, 164 (2015)
  [arXiv:1504.01692]; 
  T.~Blum {\it et al.} [RBC and UKQCD Collaborations],
  arXiv:1411.7017 [hep-lat].
\bibitem{Bruno:2014jqa} 
  M.~Bruno {\it et al.},
  JHEP {\bf 1502}, 043 (2015)
  [arXiv:1411.3982].
\bibitem{Abdel-Rehim:2015pwa} 
  A.~Abdel-Rehim {\it et al.} [ETM Collaboration],
  arXiv:1507.05068; 
  R.~Baron {\it et al.},
  JHEP {\bf 1006}, 111 (2010)
   [arXiv:1004.5284].
\bibitem{koutsou} G. Koutsou, http://www.cyprusconferences.org/einn2015/ and private communication.
\bibitem{Ishikawa:2015rho} 
  K.-I.~Ishikawa {\it et al.},
  PoS LATTICE {\bf 2015}, 075 (2015)
  [arXiv:1511.09222].
\bibitem{Alexandrou:2008tn} 
  C.~Alexandrou {\it et al.} [European Twisted Mass Collaboration],
  Phys.\ Rev.\ D {\bf 78}, 014509 (2008)
   [arXiv:0803.3190].
\bibitem{Orginos:2015aya} 
  K.~Orginos, A.~Parreno, M.~J.~Savage, S.~R.~Beane, E.~Chang and W.~Detmold,
  arXiv:1508.07583.
\bibitem{Alexandrou:2014wca} 
  C.~Alexandrou {\it et al.}
  PoS LATTICE {\bf 2014}, 151 (2015)
  [arXiv:1411.3494].
\bibitem{Anselmino:2013vqa} 
  M.~Anselmino, M.~Boglione, U.~D'Alesio, S.~Melis, F.~Murgia and A.~Prokudin,
  Phys.\ Rev.\ D {\bf 87}, 094019 (2013)
   [arXiv:1303.3822].
\bibitem{Radici:2015mwa} 
  M.~Radici, A.~Courtoy, A.~Bacchetta and M.~Guagnelli,
  JHEP {\bf 1505}, 123 (2015)
   [arXiv:1503.03495].
\bibitem{Lin:2007ap} 
  H.~W.~Lin and K.~Orginos,
  Phys.\ Rev.\ D {\bf 79}, 034507 (2009)
   [arXiv:0712.1214].
\bibitem{Green:2012ud} 
  J.~R.~Green, M.~Engelhardt, S.~Krieg, J.~W.~Negele, A.~V.~Pochinsky and S.~N.~Syritsyn,
  Phys.\ Lett.\ B {\bf 734}, 290 (2014)
   [arXiv:1209.1687];
  G.~S.~Bali {\it et al.},
  Phys.\ Rev.\ D {\bf 90}, no. 7, 074510 (2014)
   [arXiv:1408.6850];
  D.~Pleiter {\it et al.} [QCDSF/UKQCD Collaboration],
  PoS LATTICE {\bf 2010}, 153 (2010)
  [arXiv:1101.2326].
\bibitem{Wijesooriya:2005ir} 
  K.~Wijesooriya, P.~E.~Reimer and R.~J.~Holt,
  Phys.\ Rev.\ C {\bf 72}, 065203 (2005)
   [nucl-ex/0509012].
\bibitem{Alekhin:2012ig} 
  S.~Alekhin, J.~Blumlein and S.~Moch,
  Phys.\ Rev.\ D {\bf 86}, 054009 (2012)
   [arXiv:1202.2281].
 \bibitem{Alexandrou:2013tfa} 
  C.~Alexandrou, V.~Drach, K.~Hadjiyiannakou, K.~Jansen, B.~Kostrzewa and C.~Wiese,
  PoS LATTICE {\bf 2013}, 289 (2014)
  [arXiv:1311.3174].
\bibitem{Abdel-Rehim:2013wlz} 
  A.~Abdel-Rehim, C.~Alexandrou, M.~Constantinou, V.~Drach, K.~Hadjiyiannakou, K.~Jansen, G.~Koutsou and A.~Vaquero,
  Phys.\ Rev.\ D {\bf 89}, no. 3, 034501 (2014)
  [arXiv:1310.6339]; 
  Comput.\ Phys.\ Commun.\  {\bf 185}, 1370 (2014)
  [arXiv:1309.2256].
\bibitem{Abdel-Rehim:2015lha} 
  A.~Abdel-Rehim, C.~Alexandrou, M.~Constantinou, K.~Hadjiyiannakou, K.~Jansen, C.~Kallidonis, G.~Koutsou and A.~V.~Avilés-Casco,
  arXiv:1511.00433.
\bibitem{Ji:2013dva} 
  X.~Ji,
  Phys.\ Rev.\ Lett.\  {\bf 110}, 262002 (2013)
   [arXiv:1305.1539 [hep-ph]].
\bibitem{Lin:2014zya} 
  H.~W.~Lin, J.~W.~Chen, S.~D.~Cohen and X.~Ji,
  Phys.\ Rev.\ D {\bf 91}, 054510 (2015)
  [arXiv:1402.1462].
\bibitem{Alexandrou:2015rja} 
  C.~Alexandrou, K.~Cichy, V.~Drach, E.~Garcia-Ramos, K.~Hadjiyiannakou, K.~Jansen, F.~Steffens and C.~Wiese,
  Phys.\ Rev.\ D {\bf 92}, 014502 (2015)
   [arXiv:1504.07455].
\bibitem{Savage:2015eya} 
  M.~J.~Savage,
  arXiv:1510.01787;
  S.~Aoki {\it et al.} [HAL QCD Collaboration],
  PTEP {\bf 2012}, 01A105 (2012)
   [arXiv:1206.5088].
\end{thebibliography}
\end{document}